\documentclass[12pt]{revtex4}
\usepackage{amssymb}
\usepackage{amsfonts}
\usepackage{amsmath}
\usepackage{amscd}
\usepackage{graphicx}
\usepackage{cases}

\begin{document}

\title{Robust generation of logical qubit singlet states with reverse engineering and optimal control with spin qubits}

\author{Yi-Hao Kang$^{1,2}$}
\author{Zhi-Cheng Shi$^{1,2}$}
\author{Jie Song$^{3}$}
\author{Yan Xia$^{1,2,}$\footnote{E-mail: xia-208@163.com}}

\affiliation{$^{1}$Department of Physics, Fuzhou University, Fuzhou 350116, China\\
             $^{2}$Fujian Key Laboratory of Quantum Information and Quantum Optics (Fuzhou University), Fuzhou 350116, China\\
             $^{3}$Department of Physics, Harbin Institute of Technology, Harbin 150001, China}

\begin{abstract}

A protocol is proposed to generate singlet states of three logical
qubits constructed by pairs of spins. Single and multiple operations
of logical qubits are studied for the construction of an effective
Hamiltonian, with which robust control fields are derived with
invariant-based reverse engineering and optimal control. Moreover,
systematic errors are further compensated by periodic modulation for
better robustness. Furthermore, resistance to decoherence of the
protocol is also shown with numerical simulations. Therefore, the
protocol may provide useful perspectives for generations of logical
qubit entanglement in spin systems.

\end{abstract}

\maketitle

\section{Introduction}\label{sec1}

Quantum entanglement is one of the most fascinating properties of
quantum mechanics. It has been shown in the past decades that
different types of entangled states, e.g., the Bell states
\cite{BellPhysics1}, Greenberger-Horne-Zeilinger (GHZ) states
\cite{Greenberger} and W states \cite{DurPRA62} can be alternative
candidates in testing non-locality \cite{CabelloPRA65} and realizing
quantum information processing (QIP) \cite{BennettNat404}. As
entangled states have been widely applied in the field of QIP,
understanding of entanglement is gradually enhanced. Thus,
increasing interests have been recently attracted by generalized
entangled states, including high-dimensional entangled states
\cite{TureciPRB75,LGWPRA76,LWAPRA83,SXQPRA89,SCPRA93}, logical qubit
entangled states \cite{FrowisPRL106,MunroNP6,FrowisPRA85,ZLPRA92},
and macroscopic entangled state \cite{RDPRA102,CYHarxiv}.

High-dimensional entangled states are entangled states defined in
Hilbert spaces with more than two dimensions. Compared with typical
two-dimensional entangled states, high-dimensional entangled states
have shown stronger violations of local realism, larger information
capability and better security in quantum key distribution
\cite{KaszlikowskiPRL85,BourennanePRA64,BrubPRL88,CerfPRL88}. The
singlet state is an important kind of high-dimensional entangled
states. As a generalization of Bell singlet state
$(|01\rangle-|10\rangle)/\sqrt{2}$, a $D$-dimensional
($D=3,4,5,...$) singlet state encoded on $D$ qubits with
$D$-dimensional basis is defined as \cite{CabelloPRL89}
\begin{equation}\label{e0}
|S_D\rangle=\frac{1}{\sqrt{D!}}\sum\limits_\nu(-1)^{\tau(\nu)}|\nu\rangle.
\end{equation}
In Eq.~(\ref{e0}), $\nu$ is a permutation of $0,1,2,...,D-1$, and
$\tau(\nu)$ is the least times required to make transpositions of
pairs of elements for the canonical order, i.e., $0,1,2,...,D-1$.
Singlet states have shown violations of Bell inequalities
\cite{MerminPRD22}, and can be used in solving various problems
\cite{CabelloPRL89,CabelloJMO50}. These attractive features have
encouraged many researchers to study high-fidelity generations of
singlet states in different physical systems during the past few
years \cite{SHPRA95,SXQPRA85,CXPRA96,SXQPRA94,CXPRA98,SXQNJP12}.

In addition, logical qubit entangled states are entangled states
encoded on logical qubits constructed by a group of physical qubits
\cite{HMKPRL99,ShawPRA78,ZJPRL109,KapitPRL120}. Logical qubit
entangled states have potential to build up decoherence-free
subspaces against collective decoherence
\cite{WaltonPRL91,SYBSR5,ZLSR6}. When a part of physical qubits
suffer from decoherence, errors may be corrected through feedback
control or entanglement concentration and purification
\cite{HLNP15,QCQIP14,WXQIP17}. To date, various types of logical
qubit entangled states, e.g., logical Bell states \cite{ZLSR62} and
concatenated GHZ states \cite{KestingPRA88,LHNP8}, have been
investigated with a lot of advantages shown. How to realize
high-fidelity preparations of these states is now a question
attracting continuously attention.

Spin systems are promising platforms with long coherence time, good
scalability and operability \cite{XZLRMP85}, which have been
exploited in generations of typical entangled states such as the
Bell states, GHZ states and W states
\cite{PaulPRA94,StefanatosPRA99,YXTPRA97,JLPRA79,CJPRA95,JYPRA99,KYHPRA100}.
To realize scalable quantum information processing, it is essential
to expand results of protocols
\cite{PaulPRA94,StefanatosPRA99,YXTPRA97,JLPRA79,CJPRA95,JYPRA99,KYHPRA100},
and study generations of multi-dimensional entanglement. However, in
contrast to research in atomic systems
\cite{SHPRA95,SXQPRA85,CXPRA96,SXQPRA94,CXPRA98,SXQNJP12}, there are
still not many works in spin systems for generations of
high-dimensional entangled states including singlet states with more
than three qubits. One of the reasons is that a single spin has only
two natural states $|\uparrow\rangle$ and $|\downarrow\rangle$.
Considering features of logical qubits, they provide extensible way
to encode quantum information in spin systems, and may be more
robust against collective decoherence
\cite{FrowisPRL106,FrowisPRA85}. From this point, we propose a
protocol to generate logical qubit singlet states in the spin
system. We analyze the dynamics of a system with three logical
qubits constructed by pairs of spins, and derive Hamiltonians for
single- and multi-qubit operations. Based on the results, robust
control fields are derived based on invariant-based reverse
engineering
\cite{CXPRA86,ImpensPRA96,OdelinRMP91,IbanezPRA87,TorronteguiPRA89,KYHPRA101}
and optimal control
\cite{RuschhauptNJP14,DaemsPRL111,DammePRA95,DammePRA96,LXJPRA88,KYHPRA102}.
Moreover, we further reduce the influence of systematic errors by
periodic modulation. Numerical simulations show the protocol holds
better robustness than protocols with time-independent couplings and
with only reverse engineering. Influence of decoherence is also
taken into account with reported coherence time of spins
\cite{XZLRMP85}, the results demonstrate the protocol can generate
singlet states with acceptable fidelities in the presence of
decoherence. With current technology in quantum dots and nuclear
magnetic resonance (NMR) systems, physical implementation of the
protocol is feasible. The systematic errors discussed in the
protocol are also relevant to realistic physical systems. Therefore,
the protocol may be helpful to generations of singlet states in spin
systems.

\section{Lewis-Riesenfeld invariant theory}\label{sec2}

Let us first briefly introduce the Lewis-Riesenfeld invariant theory
\cite{LewisJMP10}. We consider a quantum system with Hamiltonian
$H(t)$. Introducing an invariant Hermitian operator $I(t)$,
satisfying ($\hbar=1$)
\begin{equation}\label{lr1}
\dot{I}(t)+i[H(t),I(t)]=0,
\end{equation}
an arbitrary solution $|\Psi(t)\rangle$ of time-dependent
Schr\"{o}dinger equation
$i|\dot{\Psi}(t)\rangle=H(t)|\Psi(t)\rangle$ can be expanded by
eigenstates of $I(t)$ as
\begin{equation}\label{lr2}
|\Psi(t)\rangle=\sum\limits_{k}C_ke^{i\alpha_k}|\Phi_k(t)\rangle.
\end{equation}
Here, $|\Phi_k(t)\rangle$ is the $k$-th eigenstate of $I(t)$, and
$C_k=\langle\Phi_k(0)|\Psi(0)\rangle$ denotes the corresponding
coefficient. $\alpha_k(t)$ is the Lewis-Riesenfeld phase acquired by
$|\Phi_k(t)\rangle$, obeying
\begin{equation}\label{lr3}
\dot{\alpha}_k(t)=i\langle\Phi_k(t)|\dot{\Phi}_k(t)\rangle-\langle\Phi_k(t)|H(t)|\Phi_k(t)\rangle,
\end{equation}
with $\alpha_k(0)=0$ and the initial time $t_i=0$. In practice, one
can realize construction of dynamic invariants and invariant-based
reverse engineering with the help of Lie algebra, which are briefly
introduced in Appendix B.

\section{Physical model and Hamiltonian}\label{sec3}

We now introduce the physical model for generations of singlet
states of three logical qubits. Considering a system with three
pairs of spins, which are used as three logical qubits, named $q_1$,
$q_2$ and $q_3$, respectively. The spins in pair $q_j$ ($j=1,2,3$)
are denoted by $q_{j_1}$ and $q_{j_2}$, respectively. Generally,
state of each logical qubit can be described with a Bell-state basis
as
\begin{equation}\label{e1}
\begin{aligned}
|0\rangle=\frac{1}{\sqrt{2}}(|\uparrow\uparrow\rangle+|\downarrow\downarrow\rangle),\
|1\rangle=\frac{1}{\sqrt{2}}(|\uparrow\downarrow\rangle+|\downarrow\uparrow\rangle),\\
|2\rangle=\frac{1}{\sqrt{2}}(|\uparrow\uparrow\rangle-|\downarrow\downarrow\rangle),\
|3\rangle=\frac{1}{\sqrt{2}}(|\uparrow\downarrow\rangle-|\downarrow\uparrow\rangle).
\end{aligned}
\end{equation}
The singlet state to be prepared reads
$|\Psi_s\rangle=(|012\rangle+|120\rangle+|201\rangle-|102\rangle-|210\rangle-|021\rangle)/\sqrt{6}$.
When spins $q_{j_1}$ and $q_{j_2}$ interact with each other through
Heisenberg exchange interactions with Hamiltonian
\cite{YXTPRA97,ZajacSci359,RussPRB97,VargasPRB100}
\begin{equation}\label{e2}
H_j=\sum\limits_{\jmath=x,y,z}\frac{J_{\jmath
j}}{4}\sigma_\jmath^{(q_{j_1})}\otimes\sigma_\jmath^{(q_{j_2})},
\end{equation}
we can obtain a single-qubit operation for logical qubit $q_j$ in
the Bell-state basis as
\begin{equation}\label{e3}
\begin{aligned}
H_j=\frac{1}{4}[(J_{xj}-J_{yj}+J_{zj})|0\rangle_j\langle0|+(J_{xj}+J_{yj}-J_{zj})|1\rangle_j\langle1|\\
+(-J_{xj}+J_{yj}+J_{zj})|2\rangle_j\langle2|
-(J_{xj}+J_{yj}+J_{zj})|3\rangle_j\langle3|],
\end{aligned}
\end{equation}
where $J_{\jmath j}$ ($\jmath=x,y,z$) is the strength of exchange
interaction in $\jmath$-direction. From Eq.~(\ref{e3}), we find that
$H_j$ only induces energy splits among different Bell states in the
Bell-state basis of logical qubit $q_j$. Thus, if we encode quantum
information on states $|0\rangle$, $|1\rangle$ and $|2\rangle$ in
the generation of three-logical-qubit singlet state, state
$|3\rangle$ is dynamically decoupled to the considered subspace
through $H_j$.

To obtain entanglement of logical qubits $\{q_j\}$, we also require
multi-qubit operations between each pair of logical qubits. Here, we
construct a two-qubit operation for logical qubits $q_1$ and $q_2$
as an example for constructions of multi-qubit operations. We assume
that two spins in pair $(q_{1_\iota},q_{2_{\iota'}})$
($\iota,\iota'=1,2$) interact with each other through exchange
interactions with the same strengths $\{g_\jmath(t)|\jmath=x,y,z\}$
in three different directions, such that the Hamitonian can be
written by \cite{YXTPRA97,ZajacSci359,RussPRB97,VargasPRB100}
\begin{equation}\label{e4}
H_{12}(t)=\sum\limits_{\jmath=x,y,z}\sum\limits_{\iota,\iota'=1,2}
\frac{g_\jmath(t)}{4}\sigma_\jmath^{(q_{1_\iota})}\otimes\sigma_\jmath^{(q_{2_{\iota'}})}.
\end{equation}
$H_{12}$ can described by a product Bell-state basis of $q_1$ and
$q_2$ as
\begin{equation}\label{e5}
\begin{aligned}
H_{12}(t)&=H_{ex}^{(12)}(t)+H_{dp}^{(12)}(t),&\\
H_{ex}^{(12)}(t)&=g_x(t)|01\rangle_{12}\langle10|+g_y(t)|12\rangle_{12}\langle21|
+g_z(t)|02\rangle_{12}\langle20|+\mathrm{H.c.},&\\
H_{dp}^{(12)}(t)&=g_x(t)|00\rangle_{12}\langle11|-g_y(t)|11\rangle_{12}\langle22|
+g_z(t)|00\rangle_{12}\langle22|+\mathrm{H.c.}&
\end{aligned}
\end{equation}
According to Eq.~(\ref{e5}), $H_{12}(t)$ can be divided into two
parts. $H_{ex}^{(12)}(t)$ exchanges the states of logical qubits
$q_1$ and $q_2$, while $H_{du}^{(12)}(t)$ only works when two
logical qubits in the same state in the Bell-state basis. With
similar deviations to the two-qubit operation of logical qubits
$q_1$ and $q_3$ ($q_2$ and $q_3$), we can also obtain $H_{13}(t)$
($H_{23}(t)$) with similar form of Eq.~(\ref{e5}). Consequently, if
we prepare the system in state $|012\rangle$ initially, the
Hamiltonian component
\begin{equation}\label{e6}
H_{dp}(t)=\sum\limits_{j<j'}H_{dp}^{(jj')}(t),
\end{equation}
does not work, thus the evolution can be studied in a
six-dimensional subspace spanned by
\begin{equation}\label{e7}
|\psi_1\rangle=|012\rangle,\ |\psi_2\rangle=|102\rangle, \
|\psi_3\rangle=|120\rangle,\ |\psi_4\rangle=|210\rangle,\
|\psi_5\rangle=|201\rangle, \ |\psi_6\rangle=|021\rangle.
\end{equation}
In this subspace, the singlet state $|\Psi_s\rangle$ can be
described by a simple form as
\begin{equation}\label{e8}
|\Psi_s\rangle=\frac{1}{\sqrt{6}}\sum\limits_{n=1}^6(-1)^{n+1}|\psi_n\rangle,
\end{equation}
and the multi-qubit Hamiltonian can also be simplified as
\begin{equation}\label{e9}
H_{mu}(t)=\sum\limits_{j<j'}H_{ex}^{(jj')}(t)
=g(t)(|\psi_1\rangle+|\psi_3\rangle+|\psi_5\rangle)
(\langle\psi_2|+\langle\psi_4|+\langle\psi_6|)+\mathrm{H.c.},
\end{equation}
with isotropic interactions $g_x(t)=g_y(t)=g_z(t)=g(t)$
\cite{RussPRB97,VargasPRB100}.

Since the singlet state $|\Psi_s\rangle$ is an eigenstate of the
Hamiltonian $H_{mu}(t)$, the generation of $|\Psi_s\rangle$ from
initial state $|\psi_1\rangle$ is impossible with only $H_{mu}(t)$.
Thus, assistance of the single-qubit Hamiltonian $\{H_j|j=1,2,3\}$
is important to the protocol. To construct an effective Hamiltonian
towards the target state $|\Psi_s\rangle$, we select interaction
strengths in $\{H_j\}$ as \cite{YXTPRA97}
\begin{equation}\label{e10}
J_{x1}=J_{z1}=J_{x2}=J_{y2}=J_{y3}=J_{z3}=J,\
J_{y1}=J_{z2}=J_{x3}=0.
\end{equation}
In this case, Hamiltonian of the total system in the rotation frame
of $R(t)=\exp(-i\sum_{j=1}^3H_jt)$ reads
\begin{equation}\label{e11}
H_{R}(t)=g(t)e^{-iJt}(e^{i3Jt}|\psi_1\rangle+|\psi_3\rangle+|\psi_5\rangle)
(\langle\psi_2|+\langle\psi_4|+\langle\psi_6|)+\mathrm{H.c.}
\end{equation}
By setting
\begin{equation}\label{e12}
g(t)=[2g_1(t)\sin(2Jt)+\sqrt{2}g_2(t)\sin(Jt)]/\sqrt{3},
\end{equation}
we obtain an effective Hamiltonian as
\begin{equation}\label{e13}
H_e(t)=ig_1(t)|\phi_1\rangle\langle\phi_2|-ig_2(t)|\phi_3\rangle\langle\phi_2|+\mathrm{H.c.},
\end{equation}
with
\begin{equation}\label{e14}
|\phi_1\rangle=|\psi_1\rangle,\
|\phi_2\rangle=\frac{1}{\sqrt{3}}(|\psi_2\rangle+|\psi_4\rangle+|\psi_6\rangle),\
|\phi_3\rangle=\frac{1}{\sqrt{2}}(|\psi_3\rangle+|\psi_5\rangle),
\end{equation}
under the condition $|g_1(t)|,|g_2(t)|\ll J$. In the next section,
we would show that invariant-based reverse engineering can be
exploited to construct evolution path for the generation of the
singlet state $|\Psi_s\rangle$ with the effective Hamiltonian
$H_e(t)$.

\section{Construction of evolution path via invariant-based reverse engineering}\label{sec4}

Now, let us design the control functions $g_1(t)$ and $g_2(t)$ via
invariant-based reverse engineering to obtain the singlet state
$|\Psi_s\rangle$. According to Eq.~(\ref{e13}), the effective
Hamiltonian can be decomposed as
\begin{equation}\label{e15}
H_e(t)=g_1(t)G_1+g_2(t)G_2+0\times G_3,
\end{equation}
with
\begin{equation}\label{e16}
G_1=\left[%
\begin{array}{ccc}
  0 &\  i &\  0 \\
  -i &\  0 &\  0 \\
  0 &\  0 &\  0 \\
\end{array}%
\right],\
G_2=\left[%
\begin{array}{ccc}
  0 &\  0 &\  0 \\
  0 &\  0 &\  i \\
  0 &\  -i &\  0 \\
\end{array}%
\right],\
G_3=\left[%
\begin{array}{ccc}
  0 &\  0 &\  i \\
  0 &\  0 &\  0 \\
  -i &\  0 &\  0 \\
\end{array}%
\right],
\end{equation}
being generators of special orthogonal Lie algebra so(3) obeying
commutation relations
\begin{equation}\label{e17}
[G_1,G_2]=iG_3,\ [G_2,G_3]=iG_1,\ [G_3,G_1]=iG_2.
\end{equation}
As invariants can be always constructed by generators of Lie algebra
\cite{TorronteguiPRA89}, we here consider an invariant in form of
\begin{equation}\label{e18}
I(t)=\lambda_1(t)G_1+\lambda_2(t)G_2+\lambda_3(t)G_3,
\end{equation}
with $\lambda_1(t)$, $\lambda_2(t)$ and $\lambda_3(t)$ being three
real time-dependent coefficients. Substituting the invariant $I(t)$
in Eq.~(\ref{e18}) and the effective Hamiltonian $H_e(t)$ in
Eq.~(\ref{e15}) into Eq.~(\ref{lr1}), we obtain
\begin{equation}\label{e19}
\dot{\lambda}_1=g_2\lambda_3,\ \dot{\lambda}_2=-g_1\lambda_3,\
\dot{\lambda}_3=g_1\lambda_2-g_2\lambda_1.
\end{equation}
As the rank of the three equations in Eq.~(\ref{e19}) is 2, a
constraint equation can be derived as
$\lambda_1^2+\lambda_2^2+\lambda_3^2=\mathrm{Const.}$ A simple way
to parameterize $\lambda_1(t)$, $\lambda_2(t)$ and $\lambda_3(t)$ is
using spherical coordinates for a unit sphere as
\begin{equation}\label{e20}
\lambda_1=\cos\theta\cos\beta,\ \lambda_2=\cos\theta\sin\beta,\
\lambda_3=\sin\theta.
\end{equation}
Consequently, the control functions can be reversely solved from
Eq.~(\ref{e19}) as
\begin{equation}\label{e21}
g_1(t)=\dot{\theta}\sin\beta-\dot{\beta}\cot{\theta}\cos\beta,\
g_2(t)=-\dot{\theta}\cos\beta-\dot{\beta}\cot{\theta}\sin\beta.
\end{equation}
In addition, eigenstates of the invariant $I(t)$ can be derived as
\begin{equation}\label{e22}
|\Phi_0(t)\rangle=\left[%
\begin{array}{c}
  \cos\theta\sin\beta \\
  -\sin\theta  \\
  \cos\theta\cos\beta  \\
\end{array}%
\right],\
|\Phi_+(t)\rangle=\left[%
\begin{array}{c}
  \sin\theta\sin\beta+i\cos\beta \\
  \cos\theta  \\
  \sin\theta\cos\beta-i\sin\beta  \\
\end{array}%
\right],\
|\Phi_-(t)\rangle=\left[%
\begin{array}{c}
  \sin\theta\sin\beta-i\cos\beta \\
  \cos\theta  \\
  \sin\theta\cos\beta+i\sin\beta  \\
\end{array}%
\right]\
\end{equation}
corresponding to eigenvalues 0, 1 and -1, respectively. The
Lewis-Riesenfeld phase acquired by $|\Phi_0(t)\rangle$ and
$|\Phi_\pm(t)\rangle$ respectively read
\begin{equation}\label{e23}
\alpha_0(t)=0,\ \
\alpha_\pm(t)=\int_0^t\pm2\dot{\beta}(t')\csc[\theta(t')]dt'.
\end{equation}
When we consider boundary conditions as
\begin{equation}\label{e24}
\theta(0)=0,\ \beta(0)=\pi/2,\ \theta(T)=\pi/4,\
\beta(T)=\arcsin(\sqrt{3}/3),
\end{equation}
with $T$ being the operation time, invariant-based engineering
allows us to evolve the system from the initial state
$|\Psi(0)\rangle=|\phi_1\rangle$ to the final state
$|\Psi(T)\rangle=(|\phi_1\rangle-\sqrt{3}|\phi_2\rangle+\sqrt{2}|\phi_3\rangle)/\sqrt{6}=|\Psi_s\rangle$
along the eigenstate $|\Phi_0(t)\rangle$. Therefore, the
construction of evolution path has been successfully completed via
invariant-based reverse engineering.

\section{Selections of control parameters via optimal control theory}\label{sec5}

Although the evolution path has been constructed through
invariant-based reverse engineering, the problem of selections of
control parameters has not yet been settled. Since the evolution
path $|\Phi_0(t)\rangle$ only requests constraints of parameters at
boundary, there still a lot of choices for specific expressions of
parameters. However, not all parameter selections can produce robust
fields against optional errors. Thus, how to select parameters for
high-fidelity control is a very important problem required to be
considered. Fortunately, optimal control theory provides us a
powerful tool to deal with the problem. Here, we consider that there
exists errors of control function $g(t)$ as $\delta g(t)$. In this
case, the faulty effective Hamiltonian reads
\begin{equation}\label{e25}
H_e'(t)=i(1+\delta)g_1(t)|\phi_1\rangle\langle\phi_2|
-i(1+\delta)g_2(t)|\phi_3\rangle\langle\phi_2|+\mathrm{H.c.}
\end{equation}

\begin{figure}
\scalebox{0.8}{\includegraphics[scale=0.6]{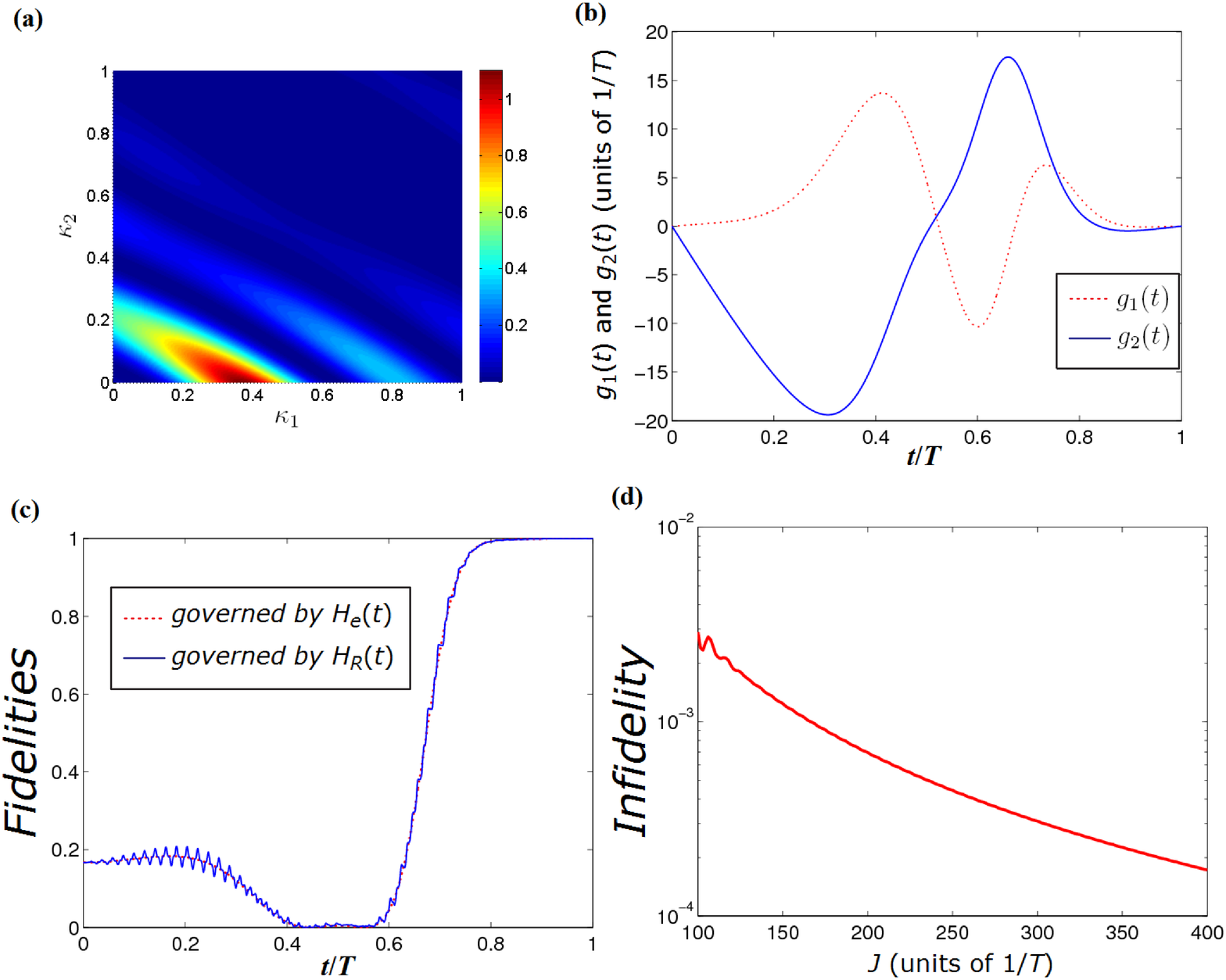}}\caption{(a)
$Q_s$ versus $\kappa_1$ and $\kappa_2$. (b) $g_1(t)$ and $g_2(t)$
versus $t/T$. (c) The fidelities
$F=|\langle\Psi(t)|\Psi_s\rangle|^2$ of obtaining the singlet state
$|\Psi_s\rangle$ versus $t/T$ with the effective Hamiltonian $H_e$
(red-dotted line) and the Hamiltonian $H_R(t)$ with $J=300/T$
(blue-solid line). (d) Infidelity $1-F$ versus $J$.}\label{fig1}
\end{figure}

To start the optimal strategy, we use time-dependent perturbation
theory \cite{RuschhauptNJP14,YXTPRA97} to derive
\begin{equation}\label{e26}
\begin{aligned}
|\Psi'(T)\rangle&\simeq|\Phi_0(T)\rangle-i\int_0^TdtU(T,t)H'(t)|\Phi_0(t)\rangle&\\
&-\int_0^Tdt\int_0^tdt'U(T,t)H'(t)U(t,t')H'(t')|\Phi_0(t')\rangle+\mathcal{O}(\delta^3),&
\end{aligned}
\end{equation}
where $|\Psi'(t)\rangle$ is state of the system with systematic
errors, $H'(t)=\delta H_e(t)$ is the total perturbation Hamiltonian,
and $U(t,t')$ reads
\begin{equation}\label{e27}
U(t,t')=\sum\limits_{k=0,\pm}e^{i[\alpha_k(t)-\alpha_k(t')]}|\Phi_k(t)\rangle\langle\Phi_k(t')|,
\end{equation}
being the evolution operator of the system in time interval $[t',t]$
without systematic errors. With the help of Eq.~(\ref{e26}), the
fidelity $F=|\langle\Psi'(T)|\Psi_s\rangle|^2$ can be estimated as
\begin{equation}\label{e28}
F\simeq1-\sum\limits_{k\neq0}|\int_0^Te^{-i\alpha_k(t)}\langle\Phi_k(t)|H'(t)|\Phi_0(t)\rangle
dt|^2,
\end{equation}
with second order of $\delta$ being kept. Defining systematic error
sensitivity as $Q_s=-\partial^2F/2\partial\delta^2$
\cite{RuschhauptNJP14}, we have
\begin{equation}\label{e30}
Q_s=2|\int_0^T(\dot{\beta}\cos\theta\sin\alpha+\dot{\theta}\cos\alpha)dt|^2,
\end{equation}
with $\alpha(t)=\alpha_+(t)$. Inspired by Ref.~\cite{DaemsPRL111},
we consider a Fourier series of $\beta(t)$ as
\begin{equation}\label{e31}
\beta(t)=\beta(T)\mu(t)+\beta(0)[1-\mu(t)]
+\kappa_1\sin[\pi\mu(t)]+\kappa_2\sin[2\pi\mu(t)],\
\mu(t)=[\theta(t)/\theta(T)]^2.
\end{equation}
Considering the interpolation of $\theta(t)$ with boundary
conditions shown in Eq.~(\ref{e24}) and
$\dot{\theta}(0)=\dot{\theta}(T)=0$ by using sine function as
\begin{equation}\label{e32}
\theta(t)=\frac{\pi}{4}\sin^2(\frac{\pi t}{2T}),
\end{equation}
we plot $Q_s$ versus $\kappa_1$ and $\kappa_2$ in
Fig.~\ref{fig1}(a), and find a local minimum for $Q_s$ at
$\kappa_1=0.97$ and $\kappa_2=0.71$.

With the parameters selected via optimal control, we plot $g_1(t)$
and $g_2(t)$ versus $t/T$ in Fig.~\ref{fig1}(b). According to
Fig.~\ref{fig1}(b), we have $g_{\max}=\max\limits_{0\leq t\leq
T}\{|g_1(t)|,|g_2(t)|\}\simeq19.4/T$. In addition, the fidelity
$F=|\langle\Psi(t)|\Psi_s\rangle|^2$ of obtaining the singlet state
$|\Psi_s\rangle$ versus $t/T$ with the effective Hamiltonian
$H_e(t)$ is plotted in Fig.~\ref{fig1}(c) by using red-dotted line.
Seen from the red-dotted line in Fig.~\ref{fig1}(c), the fidelity is
gradually approaching 1 during the evolution. This result shows that
the evolution path constructed by invariant-based reverse
engineering and parameters selected with optimal control are
successfully applied to the effective Hamiltonian. Since the
effective Hamiltonian is based on the assumption
$|g_1(t)|,|g_2(t)|\ll J$, we need to select a proper value for $J$
to make sure the effective Hamiltonian valid. Therefore, in
Fig.~\ref{fig1}(d), we plot the infidelity $1-F$ versus $J$ with the
Hamiltonian $H_R(t)$ shown in Eq.~(\ref{e11}) from $J=100/T$ to
$J=400/T$ (about fivefold to twentyfold value compared with
$g_{\max}$). According to Fig.~\ref{fig1}(d), the infidelity
decreases with the increase of coupling strength $J$ since real
dynamics of the system is more close to that governed by the
effective Hamiltonian $H_e(t)$ with a larger $J$. As an example, we
consider $J=300/T$ and plot the fidelity $F$ versus $t/T$ with the
Hamiltonian $H_R(t)$ in Fig.~\ref{fig1}(c) by using blue-solid line.
From Fig.~\ref{fig1}(c), we can see that the fidelity of obtaining
the single state is $F=0.9997$ at $t=T$.

\begin{figure}
\scalebox{0.8}{\includegraphics[scale=0.8]{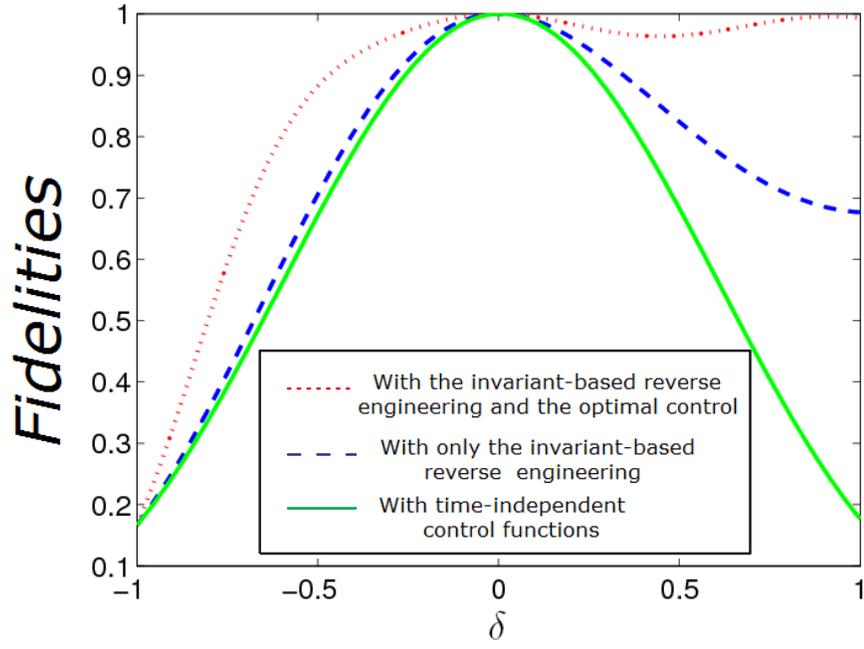}}\caption{The
fidelities of obtaining the singlet state versus $\delta$ with the
invariant-based reverse engineering and the optimal control
(red-dotted line), with only reverse engineering (blue-dashed line),
and with the time-independent control functions (green-solid
line).}\label{fig2}
\end{figure}

Now, with the coupling strengths $g(t)$ and $J$ being settled, we
now demonstrate the effect of optimal control to the generation of
the singlet state via numerical simulations. Firstly, we plot the
fidelity of obtaining the singlet state versus $\delta$ in
Fig.~\ref{fig2} by the red-dotted line. As comparisons, we plot
fidelities of obtaining the singlet state with another two group of
control functions $g_1(t)$ and $g_2(t)$. One group is using
time-independent control functions $g_1(t)=1.2358/T$ and
$g_2(t)=1.2057/T$, where the fidelity of obtaining the singlet state
versus $\delta$ is plotted in Fig.~\ref{fig2} by the blue-dashed
line. The other group is using control functions given by
invariant-based reverse engineering with parameters
\begin{equation}\label{e34}
\theta(t)=\frac{\pi}{4}\sin^2(\frac{\pi t}{2T}),\
\beta(t)=\beta(0)+[\beta(T)-\beta(0)]\sin^4(\frac{\pi t}{2T}),
\end{equation}
which are two arbitrary functions satisfying the boundary
conditions, and the fidelity of obtaining the singlet state is
plotted in Fig.~\ref{fig2} by the green-solid line. Seen from the
red-dotted line in Fig.~\ref{fig2}, we can find that the fidelity is
quite robust against systematic error of the control field $g(t)$
when it is designed by inverse engineering and optimal control. We
can find in the figure that the fidelity is kept higher than 90\% in
a very large range $\delta\in[-0.46,1]$. However, seen from
blue-dashed line, when using inverse engineering without optimal
control, the fidelity is more sensitive to the systematic error of
$g(t)$, where one can only obtain $F\geq90\%$ with
$\delta\in[-0.28,0.335]$. Furthermore, for time-independent control
functions, the situation is much worse. According to the green-solid
line in Fig.~\ref{fig2}, the range of obtaining $F\geq90\%$ is
$\delta\in[-0.25,0.26]$, even narrower than that with only inverse
engineering. Thus, optimal control help us greatly improve the
robustness against the systematic error of the control fields
$g(t)$.

\section{Further optimization by using periodic modulation}\label{sec6}

\begin{figure}
\scalebox{0.8}{\includegraphics[scale=0.58]{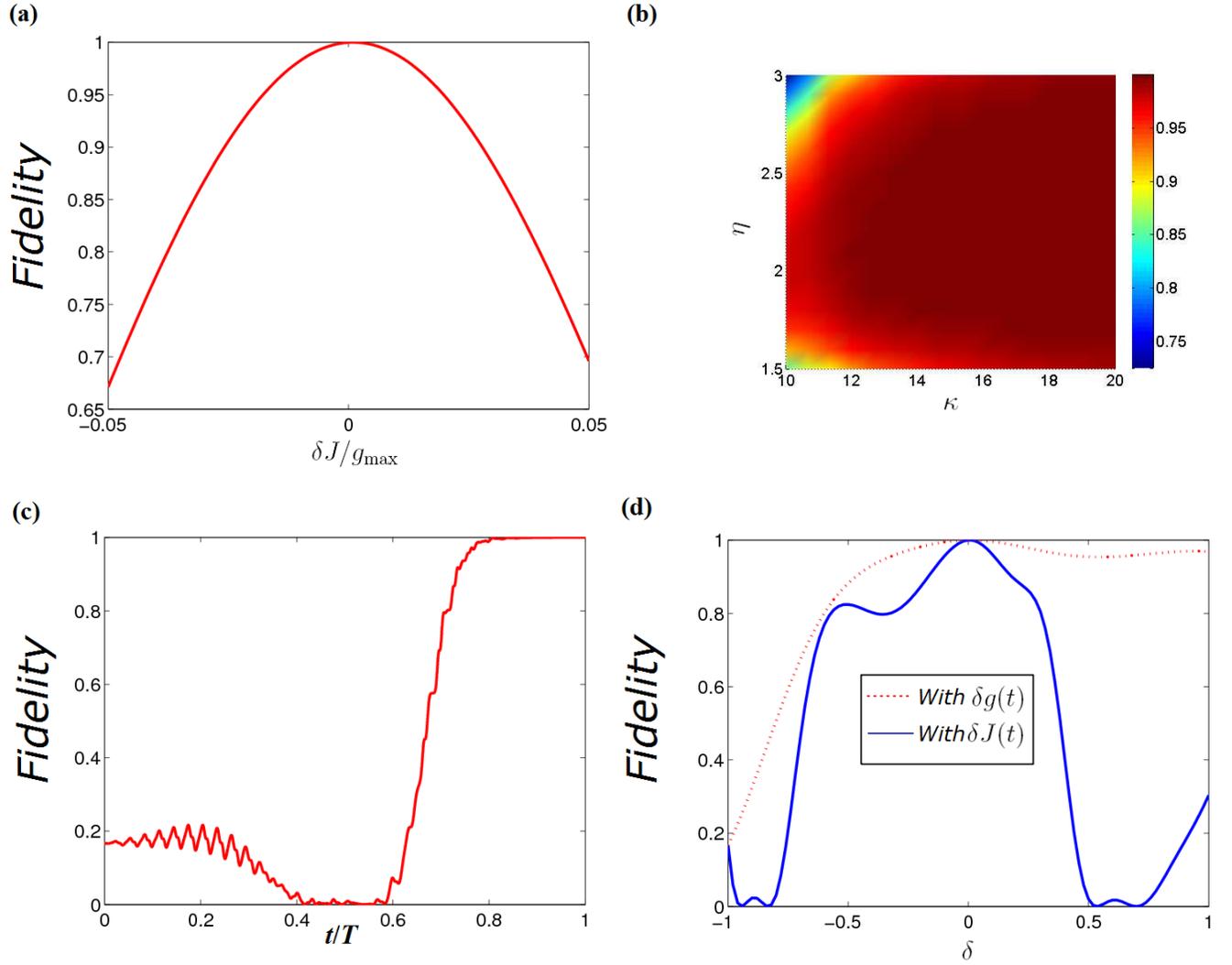}}\caption{(a)
The fidelity of obtaining the singlet state versus $\delta
J/g_{\max}$ without using the periodical modulation. (b) The
fidelity of obtaining the singlet state $|\Psi_s\rangle$ versus
$\kappa$ and $\eta$. (d) The fidelity of obtaining the singlet state
versus $\delta$ with the systematic errors $\delta g(t)$ (red-dotted
line) and $\delta J(t)$ (blue-solid line).}\label{fig3}
\end{figure}

In Sec. \ref{sec5}, we have used optimal control theory to help us
to weaken influence of the systematic error $\delta g(t)$. However,
the systematic error of the coupling strength $J$ has not yet been
considered. Since the condition $J\gg|g(t)|$ is exploited to build
up the effective Hamiltonian, the systematic error of $J$ may not be
considered as a perturbation in some cases. For example, when
$J=300/T$, ten percent of $J$ is even larger than the maximal value
of $g(t)$. This may make time-dependent perturbation theory invalid
in processing the systematic error of $J$. In Fig.~\ref{fig3}(a), we
plot fidelity of obtaining the singlet state versus $\delta
J/g_{\max}$ with $\delta J$ being the systematic error of $J$. We
can see from Fig.~\ref{fig3}(a) that, the fidelity decreases a lot
with the increase of $\delta J$. When $\delta J=-0.05g_{\max}$ and
$\delta J=0.05g_{\max}$, the fidelities are only about $67.72\%$ and
$70.19\%$, respectively. To obtain fidelities higher than 90\%, one
should control $\delta J/g_{\max}$ in $[-0.0258,0.0278]$. This may
be sometimes a strict limit for a real experiment.

According to Eq.~(\ref{e12}), we can find that $\delta J$ breaks the
resonance. When systematic error $\delta J$ appears, the faulty
effective Hamiltonian should be
\begin{equation}\label{e35}
H_e''(t)=H_e(t)+\delta
J(3|\phi_1\rangle\langle\phi_1|+|\phi_2\rangle\langle\phi_2|).
\end{equation}
The second term of Eq.~(\ref{e35}) leads rapid oscillations of
phases when $\delta J/g_{\max}$ can not be neglected, and evolution
paths given by $H_e(t)$ become no longer valid. Thus, to build up an
evolution path insensitive to $\delta J$, we should find an
alternative resonance condition independent to $J$, so that $\delta
J$ may not lead significant influence to the evolution. From this
point, we exploit another way, periodic modulation, to optimize the
fidelity under influence of $\delta J$. Here, we consider a
periodical coupling strength as \ $J(t)=J_0\cos(\omega t)$. In the
rotation frame of $R(t)=\exp(-i\int_0^tH_s(t')dt')$, the Hamiltonian
of the system becomes
\begin{equation}\label{e36}
H_R(t)=g(t)e^{-i\eta\sin(\omega t)}[e^{3i\eta\sin(\omega
t)}|\psi_1\rangle+|\psi_3\rangle+|\psi_5\rangle]
(\langle\psi_2|+\langle\psi_4|+\langle\psi_6|)+\mathrm{H.c.},
\end{equation}
with $\eta=J_0/\omega$. Considering Fourier expansions with a period
$\tilde{T}=2\pi/\omega$, we have
\begin{equation}\label{e37}
e^{-i\eta\sin(\omega
t)}=\sum\limits_{m=-\infty}^{+\infty}\mathcal{J}_m(\eta)e^{-im\omega
t},\ e^{2i\eta\sin(\omega
t)}=\sum\limits_{m=-\infty}^{+\infty}\mathcal{J}_m(2\eta)e^{im\omega
t},
\end{equation}
with the $m$-th Bessel function $\mathcal{J}_m(\eta)$. By setting
the control function $g(t)$ as
\begin{equation}\label{e38}
g(t)=[\bar{g}_1(t)\sin(\omega t)+\bar{g}_2(t)\sin(3\omega
t)]/\sqrt{3},
\end{equation}
the effective Hamiltonian $\tilde{H}_e(t)$ with periodic modulation
can be derived as
\begin{equation}\label{e39}
\begin{aligned}
&\tilde{H}_e(t)=i\tilde{g}_1(t)|\phi_1\rangle\langle\phi_2|-i\tilde{g}_2(t)|\phi_3\rangle\langle\phi_2|+\mathrm{H.c.},&\cr\cr
&\tilde{g}_1(t)=\mathcal{J}_1(2\eta)\bar{g}_1(t)+\mathcal{J}_3(2\eta)\bar{g}_2(t),&\cr\cr
&\tilde{g}_2(t)=\sqrt{2}[\mathcal{J}_1(\eta)\bar{g}_1(t)+\mathcal{J}_3(\eta)\bar{g}_2(t)],&
\end{aligned}
\end{equation}
under the condition $|\bar{g}_1(t)|,|\bar{g}_2(t)|\ll\omega$. Since
$\tilde{H}_e(t)$ has the same form as $H_e(t)$ shown in
Eq.~(\ref{e13}), the evolution governed by $\tilde{H}_e(t)$ can also
be studied by invariant-based reverse engineering. Consequently, we
can derive expressions of $\tilde{g}_1(t)$ and $\tilde{g}_2(t)$ in
the same way. Then, one can reversely obtain $g(t)$ with
\begin{equation}\label{e40}
\begin{aligned}
&\bar{g}_1(t)=[\sqrt{2}\mathcal{J}_3(\eta)\tilde{g}_1(t)-\mathcal{J}_3(2\eta)\tilde{g}_2(t)]/\Upsilon,&\cr\cr
&\bar{g}_2(t)=[-\sqrt{2}\mathcal{J}_1(\eta)\tilde{g}_1(t)+\mathcal{J}_1(2\eta)\tilde{g}_2(t)]/\Upsilon,&\cr\cr
&\Upsilon=\sqrt{2}[\mathcal{J}_1(2\eta)\mathcal{J}_3(\eta)-\mathcal{J}_3(2\eta)\mathcal{J}_1(\eta)].&
\end{aligned}
\end{equation}

With Eqs.~(\ref{e39}-\ref{e40}), let us analyze influence of the
systematic errors $\delta J(t)$ and $\delta g(t)$ again. For $\delta
g(t)$, it leads a faulty Hamiltonian as $\delta\tilde{H}_e(t)$, and
this kind of error has been optimized by optimal control in Sec.
\ref{sec5}. For $\delta J$, its leads a faulty Hamiltonian as
\begin{equation}\label{e41}
H_J'(t)=\delta J_0\cos(\omega
t)(3|\phi_1\rangle\langle\phi_1|+|\phi_2\rangle\langle\phi_2|).
\end{equation}
If we move to the frame of $R'(t)=\exp(-i\int_0^tH_J'(t')dt')$,
different from the case without periodic modulation, where $\delta
J$ leads rapid oscillations of phases, $\delta J(t)$ here only
induces small error of the parameter $\eta$ in the effective
coupling strengths $\tilde{g}_1(t)$ and $\tilde{g}_2(t)$ as
$\delta\eta=\delta J_0/\omega$ when $\delta J_0\ll\omega$.

Similar result can also be obtained in the frame of $R(t)$ without
using $R'(t)$ when $\omega\gg\delta J_0$. Firstly, we investigate
the evolution in a single period $[0,\tilde{T}]$, where the
evolution operator can be approximately described as
\begin{equation}\label{e42}
\tilde{U}_e'(\tilde{T},0)\simeq1-i\int_0^{\tilde{T}}[H_J'(t)+\tilde{H}_e(t)]dt=1-i\int_0^{\tilde{T}}\tilde{H}_e(t)dt\simeq\tilde{U}_e(\tilde{T},0),
\end{equation}
with $\tilde{U}_e'(\tilde{T},0)$ $(\tilde{U}_e(\tilde{T},0))$ being
the faulty (perfect) evolution operator in time interval
$[0,\tilde{T}]$, since $\delta J_0\tilde{T}\ll1$. Thus, the total
evolution operator (assuming $T=M\tilde{T}$, $M\in N^+$)
\begin{equation}\label{e43}
\begin{aligned}
\tilde{U}_e'(T,0)&=\tilde{U}_e'(T,(M-1)\tilde{T})...\tilde{U}_e'(2\tilde{T},\tilde{T})\tilde{U}_e'(\tilde{T},0)&\cr\cr&
\simeq\tilde{U}_e(T,(M-1)\tilde{T})...\tilde{U}_e(2\tilde{T},\tilde{T})\tilde{U}_e(\tilde{T},0)=\tilde{U}_e(\tilde{T},0),&
\end{aligned}
\end{equation}
is nearly perfect with $\omega\gg\delta J_0$. Apart from
optimization to fidelity under influence of $\delta J(t)$, periodic
modulation also bring some other benefits. For example, we have
$R(T)=1$ $(T=M\tilde{T})$ with periodic modulation, i.e., the
rotation frame of $R(t)$ coincides with the original frame at the
final time. This means we do not need any additional operation to
bring the singlet state in the rotation frame back to the original
frame. For that without periodic modulation, although we can apply
an additional operation in time interval $[T,T']$ with coupling
strengths $J\rightarrow J'$, $g(t)=0$ and $J'(T'-T)+JT=2m\pi$
($m=0,1,2,...$) to bring the singlet state back to the original
frame, more operation time is required. Thus, periodic modulation
may help to save the total interaction time if entanglement
generations are considered in the original frame.

To check validity of periodic modulation and select suitable
parameters for $\omega$ and $J_0$, we plot the fidelity of obtaining
the singlet state $|\Psi_s\rangle$ versus $\kappa=\omega T/2\pi$ and
$\eta$ in Fig.~\ref{fig3}(b). According to Fig.~\ref{fig3}(b), the
fidelity increases with the increase of $\kappa$ since the effective
Hamiltonian requires the condition $|g(t)|\ll\omega$ well satisfied.
But for values of $\eta$, the maximal fidelity appears about
$\eta=2.3$ when $\kappa$ is fixed. We find that $\eta=2.3$ is a
local maximum for $|\Upsilon|$ in Eq.~(\ref{e40}). According to
Eq.~(\ref{e40}), a larger value for $|\Upsilon|$ produces smaller
amplitude for $g(t)$. Consequently, when $\eta$ is selected for a
larger $|\Upsilon|$, the condition $|g(t)|\ll\omega$ satisfy better,
and one obtains higher fidelity. Thus, we select $\eta=2.3$ and
$\kappa=16$ in the following discussions. The fidelity of obtaining
the singlet state versus $t/T$ with periodic modulation is plotted
in Fig.~\ref{fig3}(c). Seen from Fig.~\ref{fig3}(c), infidelity of
the preparation of singlet state is only $1.377\times10^{-4}$, while
the maximal value of $J(t)$ in this case is $J_0=231.22/T$. Compared
with the result without using periodic modulation, we can obtain
higher fidelity of singlet state with smaller coupling strength
$J(t)$.

Now, let us show that the generation of the singlet state is
robustness against both systematic errors $\delta J(t)$ and $\delta
g(t)$ with numerical simulations. Firstly, we plot fidelity of
obtaining the singlet state versus $\delta$ with systematic errors
$\delta g(t)$ by the red-dotted line in Fig.~\ref{fig3}(d). Seen
from Fig.~\ref{fig3}(d), shape of the red-dotted line is similar to
that in Fig.~\ref{fig2}. The range of obtaining fidelities higher
than 90\% is also $\delta\in[-0.46,1]$. This is because the
structures of the effective Hamiltonian and the faulty Hamiltonian
leads by $\delta g(t)$ are both unchanged with periodic modulation.
As a result, the parameters selected in optimal control theory in
Sec. \ref{sec5} are also applicable with periodic modulation.
Accordingly, the robustness against $\delta g(t)$ is also reserved
when we consider periodic modulation. On the other hand, we plot the
fidelity of obtaining the singlet state versus $\delta$ with
systematic errors $\delta J(t)$ by the blue-solid line in
Fig.~\ref{fig3}(d). According to the blue-solid line in
Fig.~\ref{fig3}(d), when considering the systematic error $\delta
J(t)$, the fidelity is kept higher than 90\% if $\delta$ is
restricted in range $[-0.16,0.18]$. By substituting
$J_0\simeq231.22/T$, we have the $\max\limits_{0\leq t\leq T}|\delta
J(t)|\leq41.62/T$ with $\delta\in[-0.16,0.18]$. Considering
$\tilde{g}_{\max}=\max\limits_{0\leq t\leq
T}\{|\bar{g}_1(t)|,|\bar{g}_2(t)|\}\simeq20.29/T$, the maximal value
of $\delta J(t)$ can be almost twice as large as $\tilde{g}_{\max}$.
Compared with the result in the case without periodic modulation,
where one can only get fidelity higher than 90\% with $\delta
J/g_{\max}\in[-0.0258,0.0278]$ (or $\delta\in[-0.0017,0.0018]$),
robustness against the systematic error of the coupling strength
$J(t)$ is greatly improved by periodic modulation. Thus, periodic
modulation make the generation of the singlet state more feasible in
a real experiment.

\section{Discussions}\label{sec7}

\subsection{Experimental considerations}

We now make some experimental considerations about the
implementation of the protocol. Firstly, we use tunable exchange
interactions with strengths $J(t)$ and $g(t)$ to realize the
generation of the singlet state. In practice, electron spins in
quantum dot systems may be an alternative candidate for the
experimental implementation of the protocol. For quantum dot
systems, tunable exchange interactions can be built up between spins
in two dots \cite{ZajacSci359,RussPRB97,VargasPRB100}. Before
implementing experiments, the relationship between strengths of
exchange interactions and gate voltages can be established by
empirical formulas with some parameters being predetermined via
measurements. Accordingly, the strengths can be controlled by tuning
gate voltages in processes of experiments. For example, we consider
the formula established in Ref.~\cite{ZajacSci359}, where strength
of exchange interaction can be written by a function of gate voltage
$V_M$ as
\begin{equation}\label{add1}
J(V_M)=\mathcal{C}\frac{V_{M_0}-V_M}{(V_M-V_{M_1})^2}\exp(-\sqrt{\frac{|V_M-V_{M_0}|}{V_{on}}}),
\end{equation}
In Eq.~(\ref{add1}), $V_{M_1}$ is the voltage at which the tunneling
barrier height is zero, $V_{M_0}$ is the voltage at which the
barrier height equals the electron energy, and $V_{on}$ is the
voltage scale of the sub-exponential increase of $J$ with $V_M$, and
$\mathcal{C}$ is an overall scale factor. Form Eq.~(\ref{add1}), it
is possible to tune the coupling strengths $J(t)$ and $g(t)$ by
converting the time dependence of the coupling strengths to that of
the gate voltages. In addition, periodic modulation can be realized
by tuning gate voltage around $V_{M_0}$.

In the generation of the singlet state without periodic modulation,
considering an available coupling strength $J=20$MHz
\cite{ZajacSci359,RussPRB97,VargasPRB100}, the total interaction
time is $T=15\mu$s, according to the selection $J=300/T$. The
interaction time is much less than the coherence time of spin qubits
in quantum dots \cite{XZLRMP85}. Moreover, the tuning frequencies of
the components $g_1(t)$ and $g_2(t)$ of $g(t)$ is less than
$2J=40$MHz, which are also feasible with fast gate voltage tuning of
quantum dots \cite{ZajacSci359}.

In addition, for the generation of the singlet state with periodic
modulation, we consider an available intensity for the coupling
strength $J(t)$ as $J_0=20$MHz
\cite{ZajacSci359,RussPRB97,VargasPRB100}. The total interaction
time is $T=11.56\mu$s given by $J_0=231.22/T$. Moreover, the tuning
frequency of $J(t)$ is $\omega=8.7$MHz, and the tuning frequency of
$g(t)$ is less than $3\omega=26.1$MHz according to Eq.~(\ref{e38}).
The tuning frequencies of $g(t)$ and $J(t)$ are also feasible with
fast gate voltage tuning in quantum dot systems \cite{ZajacSci359}.
Thus, the tuning of strength is feasible in the protocol by using
the technology shown in Ref.~\cite{ZajacSci359}.

On the other hand, the systematic errors of coupling strengths
discussed in Sec. \ref{sec5} and Sec. \ref{sec6} are also proper to
describe the influence of charge noise in quantum dot systems.
Previous protocols \cite{RussPRB97,VargasPRB100} have shown that
charge noise is a main disturbing factor leading fluctuations of
electric potentials near dots, and results in deviations of coupling
strengths as $J(t)\rightarrow J(t)+\delta J(t)$ and $g(t)\rightarrow
g(t)+\delta g(t)$ in the first order approximation. Typically, the
noise level is about 1\% of the original coupling strength
\cite{VargasPRB100}. With optimizations in the protocol, fidelity of
obtaining the logical qubit singlet state can be higher than 0.998.

Apart from electron spins in quantum dot system, nuclear spins in
NMR systems may also be considered to implement the protocol. Recent
experimental protocols \cite{JYPRA99,PXPRL103} have shown tunable
exchange interactions can be effectively realized via NMR pulsing
techniques \cite{PXPRL103,VandersypenRMP76}. With the help of
external magnetic field sequences and natural
$\sigma_z\otimes\sigma_z$ couplings of nuclear spins, exchange
interactions in different directions with different strengths can be
simulated. Furthermore, the undesired interactions can be switched
off via the refocusing technique shown in
Ref.~\cite{VandersypenRMP76}. Therefore, any operations in a short
time interval can be equivalently produced by the NMR pulsing
techniques, and the total operation can be composed through
Trotter's formula \cite{SuzukiPJASB69}. The discussions about the
systematic errors of coupling strengths can also applied to reduce
influence of imperfect refocusing and deviations of coupling
strengths of nuclear spins due to imperfect calibration.

\subsection{Performance in decoherence environment}

\begin{figure}
\scalebox{0.8}{\includegraphics[scale=0.8]{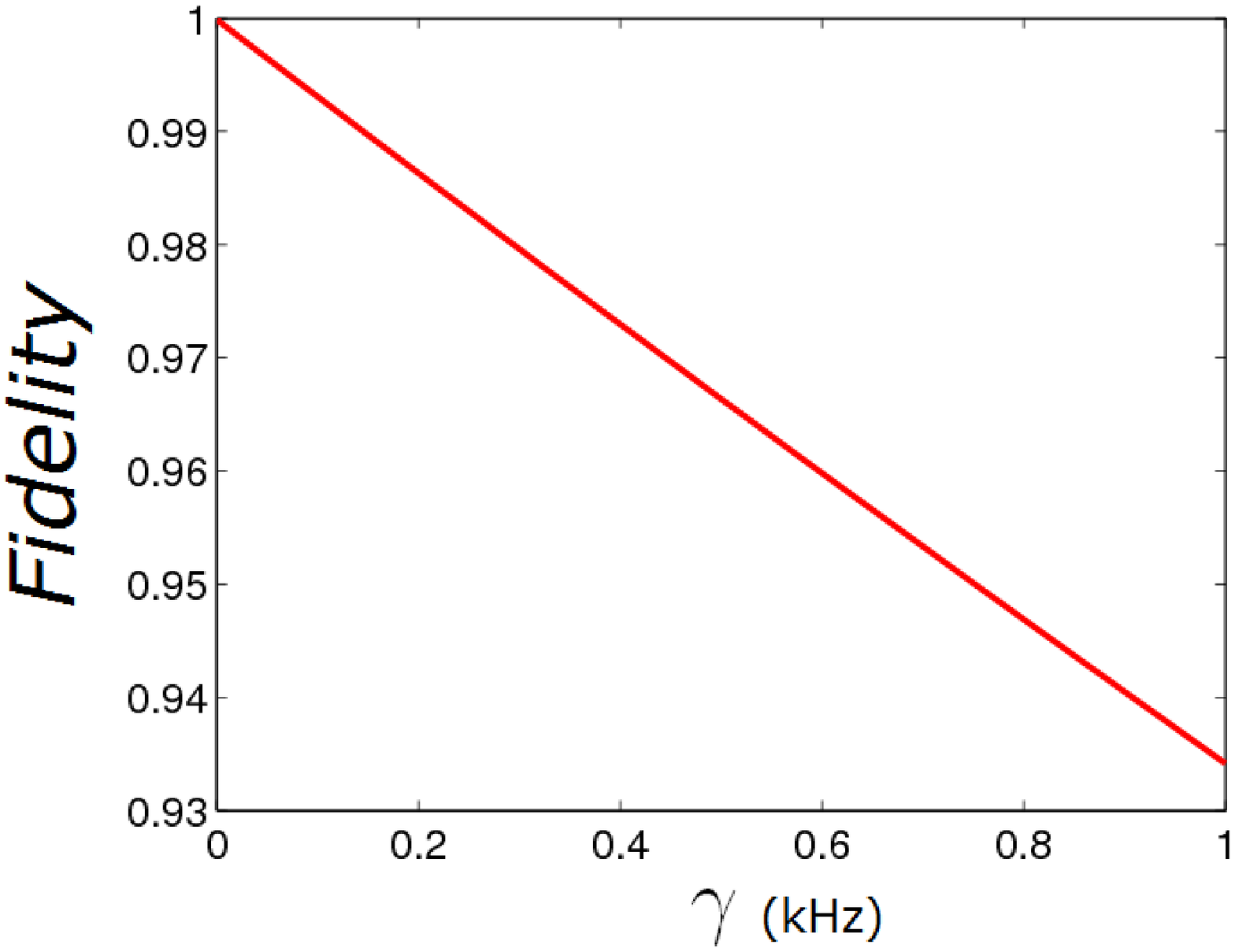}}\caption{Fidelities
of obtaining the singlet state versus dephasing rate
$\gamma$.}\label{fig4}
\end{figure}

When a spin system is not well isolated from environment,
decoherence may disturb the unitary evolution of the spin systems.
Ref.~\cite{BayatNJP17} have shown that the dephasing due to random
energy fluctuations induced on qubit levels by random magnetic and
electric fields in the environment, is one of the challenges in real
experiments when the number of spins increases. Under influence of
dephasing, evolution of the system is governed by the master
equation \cite{BayatNJP17}
\begin{equation}\label{disa1}
\dot{\rho}(t)=i[\rho(t),H(t)]
+\sum\limits_{j=1}^3\sum\limits_{\iota=1}^2\gamma[\sigma_{z}^{(q_{j_\iota})}\rho(t)\sigma_{z}^{(q_{j_\iota})}
-\rho(t)],
\end{equation}
where $\gamma$ is the dephasing rate. Based on Eq.~(\ref{disa1}),
the fidelity of obtaining the singlet state versus dephasing rate
$\gamma$ is plotted in Fig.~\ref{fig4}. Considering reported
coherence time of electron spins in quantum dots about ms-s
\cite{XZLRMP85}, the range of $\gamma$-axis is set from 0 to 1kHz.
As shown by Fig.~\ref{fig4}, the fidelity is 0.9931 when
$\gamma=0.1$kHz. The protocol can still perform well when the
coherence time is about 10ms. Moreover, for $\gamma=1$kHz, the
fidelity is still 0.9342. Therefore, even when we consider a shorter
coherence time about 1 ms, the protocol can still produce fidelities
higher than 0.9.

\section{Conclusion}

In conclusion, we have proposed a protocol to realize the generation
of singlet states of three logical qubits constructed by pairs of
spins. The protocol contains comprehensive consideration of control
Hamiltonian, invariant-based reverse engineering, optimal control
and periodic modulation. Firstly, single-qubit and multi-qubit
interactions of logical qubits were analyzed, and the effective
Hamiltonian was built up. Secondly, invariant-based reverse
engineering was successfully used in studying the evolution governed
by the effective Hamiltonian, and an evolution path was established
for the generation of the target state. Thirdly, optimal control
theory was exploited in selections of control parameters. We showed
that the entanglement generation is robust against the systematic
error of coupling strength $g(t)$. Fourthly, to further improve
robustness of the protocol, we applied periodic modulation to
coupling strengths $g(t)$ and $J(t)$, and derived an effective
Hamiltonian with similar structure as that in the case without
periodic modulation. In this way, the preparation of the singlet
state benefited from invariant-based reverse engineering, optimal
control and periodic modulation simultaneously, and became robust
against systematic errors of coupling strengths $g(t)$ and $J(t)$.
As the protocol can produce acceptable fidelity in a relative wide
range of systematic errors, and spin systems also possess nice
inherent controllability
\cite{ZajacSci359,RussPRB97,PXPRL103,VargasPRB100,VandersypenRMP76}
to modulate time-dependent couplings in tolerable error ranges, the
protocol may be feasible in experiments. We hope the protocol can be
helpful to robust generations of logical qubit singlet state with
spin systems.

\section*{Acknowledgements}

This work was supported by the National Natural Science Foundation
of China under Grant No. 11805036.

\section*{Appendix: Initialization of logical qubits}

Since quantum information of logical qubit is encoded in Bell-state
basis $|0\rangle$, $|1\rangle$ and $|2\rangle$, and the initial
state we considered in the generation of singlet state is
$|012\rangle$, discussions about initialization of logical qubits
may be useful. Here, Hamiltonian for initialization of the logical
qubit $q_j$ is considered as
\begin{equation}\label{e45}
\begin{aligned}
H_{i}(t)&=\frac{B_x(t)}{2}[\sigma_x^{(q_{j_1})}+\sigma_x^{(q_{j_1})}]
+\frac{B_z(t)}{2}[\sigma_z^{(q_{j_1})}+\sigma_z^{(q_{j_1})}]&\cr\cr
&+\frac{J_{yj}}{4}\sigma_y^{(q_{j_1})}\otimes\sigma_y^{(q_{j_2})},&
\end{aligned}
\end{equation}
where $B_x(t)$ and $B_z(t)$ are magnetic fields along $x$-axis and
$z$-axis. In the Bell-state basis
$\{|0\rangle_j,|1\rangle_j,|2\rangle_j,|3\rangle_j\}$, $H_{i}(t)$
can be described as
\begin{equation}\label{e46}
\begin{aligned}
&H_{in}(t)=H_{inB}(t)+H_{inJ},&\cr\cr
&H_{inB}(t)=[B_x(t)|0\rangle\langle1|+B_z(t)|0\rangle\langle2|]/2+\mathrm{H.c.},&\cr\cr
&H_{inJ}=\frac{J_{yj}}{4}(|1\rangle_j\langle1|+|2\rangle_j\langle2|-|0\rangle_j\langle0|-|3\rangle_j\langle3|).&
\end{aligned}
\end{equation}
In the rotation frame of $R_{in}(t)=\exp(-iH_{inJ}t)$, the
Hamiltonian becomes
\begin{equation}\label{e47}
H_{inR}(t)=e^{-iJ_{yj}t/2}[B_x(t)|0\rangle\langle1|+B_z(t)|0\rangle\langle2|]/2+\mathrm{H.c.}
\end{equation}
If we set the variations of magnetic fields as
\begin{equation}\label{e48}
B_x(t)=2\bar{B}_x(t)\sin(J_{yj}t),\
B_z(t)=-2\bar{B}_z(t)\sin(J_{yj}t),\
\end{equation}
we can obtain an effective Hamiltonian as
\begin{equation}\label{e49}
H_{ine}(t)=i\bar{B}_x(t)|1\rangle\langle0|-i\bar{B}_z(t)|2\rangle\langle0|+\mathrm{H.c.}
\end{equation}
Noticing that $H_{ine}(t)$ possesses the same structure as $H_e(t)$
in Eq.~(\ref{e13}) with $\bar{B}_x(t)\rightarrow g_1(t)$ and
$\bar{B}_z(t)\rightarrow g_2(t)$, the invariant $I(t)$ in form of
Eq.~(\ref{e18}) can still be applied to $H_{ine}(t)$ by changing
basis from $\{|\phi_1\rangle,|\phi_2\rangle,|\phi_3\rangle\}$ to the
basis $\{|1\rangle,|0\rangle,|2\rangle\}$, and the eigenstate
$|\Phi_0(t)\rangle$ of $I(t)$ can still be used as an evolution path
for initialization. Assuming that the initial state is the product
state
$|\downarrow\downarrow\rangle_{q_{j1}q_{j2}}=\frac{1}{\sqrt{2}}(|0\rangle_j-|2\rangle_j)$,
the system can evolves from
$|\downarrow\downarrow\rangle_{q_{j1}q_{j2}}$ to $|0\rangle_j$
($|1\rangle_j$, $|2\rangle_j$) when the boundary condition is set as
$\{\theta(0),\beta(0)\}=\{-3\pi/4,0\}$ and
$\{\theta(T),\beta(T)\}=\{-\pi/2,\mathrm{arbitrary\ angle}\}$
($\{\theta(T),\beta(T)\}=\{0,\pi/2\}$,
$\{\theta(T),\beta(T)\}=\{0,0\}$).

Moreover, initialization of logical qubits can also optimized by
optimal control and periodic modulation. For example, to use
periodic modulation, one can set
\begin{equation}\label{e50}
\begin{aligned}
&J_{yj}(t)=2J_{y0}\cos(\omega t),&\cr\cr
&B_x(t)=2\tilde{B}_x(t)\sin(\omega
t)/\mathcal{J}_1(J_{y0}/\omega),&\cr\cr
&B_z(t)=-2\tilde{B}_z(t)\sin(\omega
t)/\mathcal{J}_1(J_{y0}/\omega).&
\end{aligned}
\end{equation}
Then, by moving into the rotation frame of
$\tilde{R}_{in}(t)=\exp{-i\int_0^tH_{inJ}(t')dt'}$, an effective
Hamiltonian can be derived as
\begin{equation}\label{e51}
\tilde{H}_{ine}(t)=i\tilde{B}_x(t)|1\rangle\langle0|-i\tilde{B}_z(t)|2\rangle\langle0|+\mathrm{H.c.}
\end{equation}
Subsequently, evolutions governed by effective Hamiltonian
$\tilde{H}_{ine}(t)$ can be further studied with invariant-based
reverse engineering and optimal control.

\newpage

\begin{thebibliography}{999}
\bibitem{BellPhysics1}J.~S.~Bell, \emph{Physics} \textbf{1965}, 1, 195.
\bibitem{Greenberger}D.~M.~Greenberger, M.~A.~Horne, A.~Shimony, A.~Zeilinger, \emph{Am. J. Phys.} \textbf{1990}, 58, 1131.
\bibitem{DurPRA62}W.~D\"{u}r, G.~Vidal, J.~I.~Cirac, \emph{Phys. Rev. A} \textbf{2000}, 62, 062314.
\bibitem{CabelloPRA65}A.~Cabello, \emph{Phys. Rev. A} \textbf{2002}, 65, 032108.
\bibitem{BennettNat404}C.~H.~Bennett, D.~P.~DiVincenzo, \emph{Nature} \textbf{2000}, 404, 247.
\bibitem{TureciPRB75}H.~E.~T\"{u}reci, J.~M.~Taylor, A.~Imamoglu, \emph{Phys. Rev. B} \textbf{2007}, 75, 235313.
\bibitem{LGWPRA76}G.~W.~Lin, M.~Y.~Ye, L.~B.~Chen, Q.~H.~Du, X.~M.~Lin, \emph{Phys. Rev. A}  \textbf{2007}, 76, 014308.
\bibitem{LWAPRA83}W.~A.~Li, G.~Y.~Huang, \emph{Phys. Rev. A} \textbf{2011}, 83, 022322.
\bibitem{SXQPRA89}X.~Q.~Shao, T.~Y.~Zheng, C.~H.~Oh, S.~Zhang, \emph{Phys. Rev. A} \textbf{2014}, 89, 012319.
\bibitem{SCPRA93}C.~Song, S.~L.~Su, J.~L.~Wu, D.~Y.~Wang, X.~Ji, S.~Zhang, \emph{Phys. Rev. A} \textbf{2016}, 93, 062321.
\bibitem{FrowisPRL106}F.~Fr\"{o}wis, W. D\"{u}r, \emph{Phys. Rev. Lett.} \textbf{2011}, 106, 110402.
\bibitem{MunroNP6}W.~J.~Munro, A.~M.~Stephens, S.~J.~Devitt, K.~A.~Harrison, K.~Nemoto, \emph{Nat. Photon.} \textbf{2012}, 6, 777.
\bibitem{FrowisPRA85}F.~Fr\"{o}wis, W. D\"{u}r, \emph{Phys. Rev. A} \textbf{2012}, 85, 052329.
\bibitem{ZLPRA92}L.~Zhou, Y.~B.~Sheng, \emph{Phys. Rev. A} \textbf{2015}, 92, 042314.
\bibitem{RDPRA102}D.~Ran, W.~J.~Shan, Z.~C.~Shi, Z.~B.~Yang, J.~Song, and Y.~ Xia, \emph{Phys. Rev. A} \textbf{2020}, 102, 022603.
\bibitem{CYHarxiv}Y.~H.~Chen, W.~Qin, X.~Wang, A.~Miranowicz, F.~Nori, arXiv:2008.04078. %https://arxiv.org/abs/2008.04078
\bibitem{KaszlikowskiPRL85}D.~Kaszlikowski, P.~Gnacinski, M.~Zukowski, W.~Miklaszewski, A.~Zeilinger, \emph{Phys. Rev. Lett.} \textbf{2000}, 85, 4418.
\bibitem{BourennanePRA64}M.~Bourennane, A.~Karlsson, G.~Bj\"{o}rk, \emph{Phys. Rev. A} \textbf{2001}, 64, 012306.
\bibitem{BrubPRL88}D. Bru{\ss}, C.~Macchiavello, \emph{Phys. Rev. Lett.} \textbf{2002}, 88, 127901.
\bibitem{CerfPRL88}N.~J.~Cerf, M.~Bourennane, A.~Karlsson, N.~Gisin, \emph{Phys. Rev. Lett.} \textbf{2002}, 88, 127902.
\bibitem{CabelloPRL89}A.~Cabello, \emph{Phys. Rev. Lett.} \textbf{2002}, 89, 100402.
\bibitem{MerminPRD22}N.~D.~Mermin, \emph{Phys. Rev. D} \textbf{1980}, 22, 356.
\bibitem{CabelloJMO50}A.~Cabello, \emph{J. Mod. Opt.} \textbf{2003}, 50, 1049.
\bibitem{SHPRA95}H.~Sun, P.~Xu, H.~Pu, W.~Zhang, \emph{Phys. Rev. A} \textbf{2017}, 95, 063624.
\bibitem{SXQPRA85}X.~Q.~Shao, T.~Y.~Zheng, S.~Zhang, \emph{Phys. Rev. A} \textbf{2012} 85, 042308.
\bibitem{CXPRA96}X.~Chen, H.~Xie, G.~W.~Lin, X.~Shang, M.~Y.~Ye, X.~M.~Lin, \emph{Phys. Rev. A} \textbf{2017}, 96, 042308.
\bibitem{SXQPRA94}X.~Q.~Shao, Z.~H.~Wang, H.~D.~Liu, X.~X.~Yi, \emph{Phys. Rev. A} \textbf{2016} 94, 032307.
\bibitem{CXPRA98}X.~Chen, G.~W.~Lin, H.~Xie, X.~Shang, M.~Y.~Ye, X.~M.~Lin, \emph{Phys. Rev. A} \textbf{2018}, 98, 042335.
\bibitem{SXQNJP12}X.~Q.~Shao, H.~F.~Wang, L.~Chen, S.~Zhang, Y.~F.~Zhao, K.~H.~Yeon, \emph{New J. Phys.} \textbf{2010}, 12, 023040.
\bibitem{HMKPRL99}M.~K.~Henry, C.~Ramanathan, J.~S.~Hodges, C.~A.~Ryan, M.~J.~Ditty, R.~Laflamme, D.~G.~Cory, \emph{Phys. Rev. Lett.} \textbf{2007}, 99, 220501.
\bibitem{ShawPRA78}B.~Shaw, M.~M.~Wilde, O.~Oreshkov, I.~Kremsky, D.~A.~Lidar, \emph{Phys. Rev. A} \textbf{2008}, 78, 012337.
\bibitem{ZJPRL109}J.~Zhang, R.~Laflamme, D.~Suter, \emph{Phys. Rev. Lett.} \textbf{2012}, 109, 100503.
\bibitem{KapitPRL120}E.~Kapit, \emph{Phys. Rev. Lett.} \textbf{2018}, 120, 050503.
\bibitem{WaltonPRL91}Z.~D.~Walton, A.~F.~Abouraddy, A.~V.~Sergienko, B.~E.~A.~Saleh, M.~C.~Teich, \emph{Phys. Rev. Lett.} \textbf{2003}, 91, 087901.
\bibitem{SYBSR5}Y.~B.~Sheng, L.~Zhou, \emph{Sci. Rep.} \textbf{2015}, 5, 13453.
\bibitem{ZLSR6}L.~Zhou, Y.~B.~Sheng, \emph{Sci. Rep.} \textbf{2016}, 6, 28813.
\bibitem{HLNP15}L.~Hu, Y.~Ma, W.~Cai, X.~Mu, Y.~Xu, W.~Wang, Y.~Wu, H.~Wang, Y.~P.~Song, C.~L.~Zou, S.~M.~Girvin, L.~M.~Duan, L.~Sun, Nat. Phys. \textbf{2019}, 15, 503.
\bibitem{QCQIP14}C.~Qu, L.~Zhou, Y.~B.~Sheng, \emph{Quantum Inf. Process.} \textbf{2015}, 14, 4131.
\bibitem{WXQIP17}X.~Wu, L.~Zhou, W.~Zhong, Y.~B.~Sheng, \emph{Quantum Inf Process.} \textbf{2018}, 17, 255.
\bibitem{ZLSR62}L.~Zhou, Y.~B.~Sheng, \emph{Sci. Rep.} \textbf{2016}, 6, 20901.
\bibitem{KestingPRA88}F.~Kesting, F.~Fr\"{o}wis, W.~D\"{u}r, \emph{Phys. Rev. A} \textbf{2013}, 88, 042305.
\bibitem{LHNP8}H.~Lu, L.~K.~Chen, C.~Liu, P.~Xu, X.~C.~Yao, L.~Li, N.~L.~Liu, B.~Zhao, Y.~A.~Chen, J.~W.~Pan, \emph{Nat. Photon.} \textbf{2014}, 8, 364.

\bibitem{XZLRMP85}Z.~L.~Xiang, S.~Ashhab, J.~Q.~You, F.~Nori, \emph{Rev. Mod. Phys.} \textbf{2013}, 85, 623.
\bibitem{PaulPRA94}K.~Paul, A.~ K.~Sarma, \emph{Phys. Rev. A} \textbf{2016}, 94, 052303.
\bibitem{StefanatosPRA99}D.~Stefanatos, E.~Paspalakis, \emph{Phys. Rev. A} \textbf{2019}, 99, 022327.
\bibitem{YXTPRA97}X.~T.~Yu, Q.~Zhang, Y.~Ban, X.~Chen,  \emph{Phys. Rev. A} \textbf{2018}, 97, 062317.
\bibitem{JLPRA79}L.~Jin, Z.~Song, \emph{Phys. Rev. A} \textbf{2009}, 79, 042341.
\bibitem{CJPRA95}J.~Chen, H.~Zhou, C.~Duan, X.~Peng, \emph{Phys. Rev. A} \textbf{2017}, 95, 032340.
\bibitem{JYPRA99}Y.~Ji, J.~Bian, X.~Chen, J.~Li, X.~Nie, H.~Zhou, X.~Peng, \emph{Phys. Rev. A} \textbf{2019}, 99, 032323.
\bibitem{KYHPRA100}Y.~H.~Kang, Z.~C.~Shi, B.~H.~Huang, J.~Song, Y.~Xia, \emph{Phys. Rev. A} \textbf{2019}, 100, 012332.

\bibitem{CXPRA86}X.~Chen, J.~G.~Muga, \emph{Phys. Rev. A} \textbf{2012}, 86, 033405.
\bibitem{ImpensPRA96}F.~Impens, D.~Gu\'{e}ry-Odelin, \emph{Phys. Rev. A} \textbf{2017}, 96, 043609.
\bibitem{OdelinRMP91}D.~Gu\'{e}ry-Odelin, A.~Ruschhaupt, A.~Kiely, E.~Torrontegui, S.~Mart\'{\i}nez-Garaot, J.~G.~Muga, \emph{Rev. Mod. Phys.} \textbf{2019}, 91, 045001.
\bibitem{IbanezPRA87}S.~Ib\'{a}\~{n}ez, X.~Chen, J.~G.~Muga, \emph{Phys. Rev. A} \textbf{2013}, 87, 043402.
\bibitem{TorronteguiPRA89}E.~Torrontegui, S.~Mart\'{\i}nez-Garaot, J.~G.~Muga, \emph{Phys. Rev. A} \textbf{2014}, 89, 043408.
\bibitem{KYHPRA101}Y.~H.~Kang, Z.~C.~Shi, B.~H.~Huang, J.~Song, Y. Xia, \emph{Phys. Rev. A} \textbf{2020}, 101, 032322.
\bibitem{RuschhauptNJP14}A.~Ruschhaupt, X.~Chen, D.~Alonso, J.~G.~Muga, \emph{New J. Phys.} \textbf{2012}, 14, 093040.
\bibitem{DaemsPRL111}D.~Daems, A.~Ruschhaupt, D.~Sugny, S.~Gu\'{e}rin, \emph{Phys. Rev. Lett.} \textbf{2013}, 111, 050404.
\bibitem{DammePRA95}L.~Van Damme, Q.~Ansel, S.~J.~Glaser, D.~Sugny, \emph{Phys. Rev. A} \textbf{2017}, 95, 063403.
\bibitem{DammePRA96}L.~Van-Damme, D.~Schraft, G.~T.~Genov, D.~Sugny, T.~Halfmann, S.~Gu\'{e}rin, \emph{Phys. Rev. A} \textbf{2017}, 96, 022309.
\bibitem{LXJPRA88}X.~J.~Lu, X.~Chen, A.~Ruschhaupt, D.~Alonso, S.~Gu\'{e}rin, J.~G.~Muga, \emph{Phys. Rev. A} \textbf{2013}, 88, 033406.
\bibitem{KYHPRA102}Y.~H.~Kang, Z.~C.~Shi, J.~Song, Y. Xia, \emph{Phys. Rev. A} \textbf{2020}, 102, 022617.

\bibitem{LewisJMP10}H.~R.~Lewis, W.~B.~Riesenfeld, \emph{J. Math. Phys.} \textbf{1969}, 10, 1458.
\bibitem{ZajacSci359}D.~M.~Zajac, A.~J.~Sigillito, M.~Russ, F.~Borjans, J.~M.~Taylor, G.~Burkard, J.~R.~Petta, \emph{Science} \textbf{2018}, 359, 439.
\bibitem{RussPRB97}M.~Russ, D.~M.~Zajac, A.~J.~Sigillito, F.~Borjans, J.~M.~Taylor, J.~R.~Petta, G.~Burkard, \emph{Phys. Rev. B} \textbf{2018}, 97, 085421.
\bibitem{VargasPRB100}F.~A.~Calderon-Vargas, G.~S.~Barron, X.~H.~Deng, A.~J.~Sigillito, E.~Barnes, S.~E.~Economou, \emph{Phys. Rev. B} \textbf{2019}, 100, 035304.
\bibitem{PXPRL103}X.~Peng, J.~Zhang, J.~Du, D.~Suter, \emph{Phys. Rev. Lett.} \textbf{2009}, 103, 140501.
\bibitem{VandersypenRMP76}L.~M.~K.~Vandersypen, I.~L.~Chuang, \emph{Rev. Mod. Phys.} \textbf{2005}, 76, 1037.
\bibitem{SuzukiPJASB69}M.~Suzuki, \emph{Proc. Jpn. Acad. Ser. B} \textbf{1993}, 69, 161.
\bibitem{BayatNJP17}A.~Bayat, Y.~Omar, \emph{New J. Phys.} \textbf{2015}, 17, 103041.

\end{thebibliography}
\end{document}